# Stochastic Variations in Nanoscale HZO based Ferroelectric finFETs: A Synergistic Approach of READ Optimization and Hybrid Precision Mixed Signal WRITE Operation to Mitigate the Implications on DNN Applications

Sourav De[1*], Md. Aftab Baig[1], Bo-Han Qiu[1], Hoang-Hiep Le[1], Po-Jung Sung[2], Chun-Jung Su[2], Yao-Jen Lee[2*], Darsen Lu[1*]
[1]Institute of Microelectronics, National Cheng Kung University, Tainan, Taiwan
[2]Taiwan Semiconductor Research Institute, Hsinchu, Taiwan.
email: ( Q18077502@gs.ncku.edu.tw, darsenlu@ncku.edu.tw, yjlee@narlabs.org.tw )

*Abstract*— This paper reports a synergistic approach of *"READ"* and *"WRITE"* optimization by deploying a high-precision digital computation unit along with a low-precision ferroelectric finFET (Fe-finFETs) based analog vector-matrix multiplication block for mitigating the impact of stochastic device variations in hafnium zirconium oxide (HZO) based Fe-finFETs. Fe-finFET devices with minimum gate length of 40 nm and fin width of 20 nm have been fabricated on silicon-on-insulator wafers using a gate-first self-align process. Device-to-device and cycle-to-cycle variation modeling is performed on the basis of measured data and has been applied to system-level neural network simulations using the CIMulator software platform. We observed that the device-to-device variation has been mostly compensated through hybrid-precision-mixed-signal (HPMS) online training of neural networks and exhibits no effect on inference accuracy, whereas cycle-to-cycle threshold voltage variation can be tolerated up to 400 mV for MNIST handwritten digit recognition. An online training accuracy of 96.34% has been achieved, given the measured variability.

## I. Introduction

The CMOS compatibility of HZO and the recent advents in the research of this material renders HZO-based Fe-finFETs an excellent candidate for logic [1], memory, and neuromorphic devices. This tendency can be attributed to their endurance and write speed being superior to flash technology, considerably higher on-to-off current ratio than magneto-resistive random-access memory (MRAM) and lesser vulnerability towards random telegraphic noise because of charge-based operations unlike resistive random-access memory (RRAM) [1-5]. However, the major problems are the innate stochastic characteristic because of the random phase distribution in the HZO crystal and the abundance of defect sites acting as possible charge trapping sites, that may capture electrons of holes from the channel side (CS) or gate side (GS) [6]. These effects induce severe reliability problem in terms of degradation of endurance [5], [6] and variation of post-programming channel conductance among devices of same geometry. These non-desiring effects becomes more severe in deeply scaled HZO-based ferroelectric FET (FE-FET) devices.

Deep neural network (DNN) consists of multiple layers of entwined neurons and during the training process the appropriate weights for the synapses are searched in order to perform the classification task without further training.

The pivotal predicament for designing Fe-finFET based DNN lies in the inability to change the conductance state of deeply scaled Fe-FET devices in a definitive way during the training process. In this work, we introduce a hardware-compatible HPMS training approach, which is similar to the $\Delta - \Sigma$ modulation method, to abate the impact of device-to-device (D2D) variation in HZO based Fe-finFETs for neuromorphic computing.

This paper begins with a discussion on the characterization and analysis of Fe-finFETs. In the second part of this paper, we evaluate the effects of experimentally observed D2D and cycle-to-cycle (C2C) variations on neuromorphic systems. We conclude that, by deploying HPMS training method the NN shows considerable immunity to low precision and hence device variations. However, the method necessitates high endurance above $10^8$ cycles and another additional high precision memory blocks to store the residue portion of the weights.

## II. Characterization of Fe-finFETs

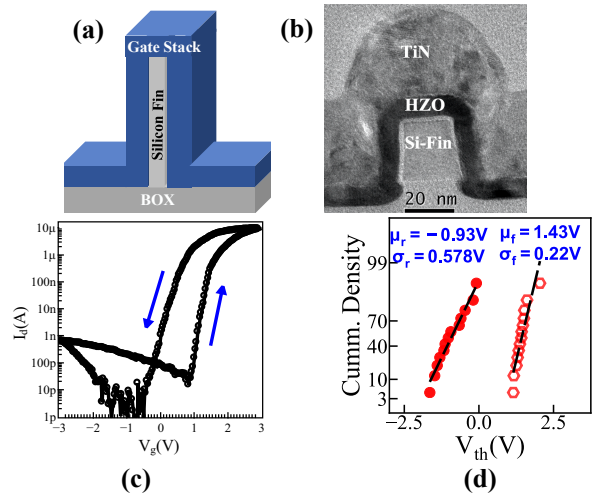

Fig. 1 (a). The schematic of the fabricated device. (b) The TEM micrograph depicts the device dimensions. (c). The $I_d$-$V_g$ curve for Fe-finFET shows CCW or ferroelectric dominant swing. (d). The distribution of forward and reverse threshold voltage shows device to device variations.

We commenced our research by fabricating the nanoscale Fe-finFETs with minimum gate length ($L_g$) of 50 nm and fin width ($T_{fin}$) of 20 nm. The process flow for fabrication has been described in our previous publications [7]. Figure. 1(a). shows the schematic of the fabricated device and Figure. 1(b). shows the transmission electron microscopic

(TEM) image of the fabricated devices. Figure. 1(c) displays the DC $I_d$-$V_g$ characteristics, where the presence of ferroelectricity has been displayed by the counterclockwise (CCW) hysteresis. The D2D variation of forward and reverse threshold voltage has been shown in Figure. 1(d). In the next sections, we discuss the analog program-erase response of deeply scaled Fe-finFET devices, which will be followed by evaluating the impact of read voltage ($V_g^{read}$) on the ON-state to OFF-state channel conductance ($G_{ch}$) ratio ($G_{ch}^{ON}/G_{ch}^{OFF}$).

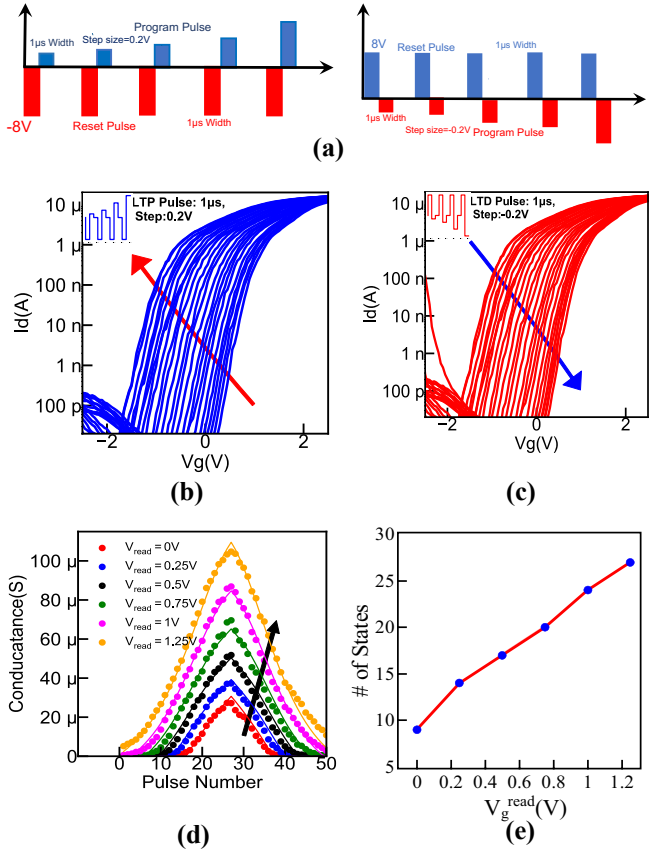

Fig. 2(a). Pulse scheme for obtaining LTP-LTD characteristics. Each programming pulse was preceeded by an erase pulse. (b). LTP characteristics using 1µs-wide pulses from 3V to 8V with 0.2V steps. ($L_g$=50nm) (c). LTD characteristics using 1µs-wide pulses from -0.8V to -5.4V with -0.2V steps. ($L_g$=50nm). (d). Characterization and modeling the change of $G_{ch}$ during LTP-LTD operation. (e). Read voltage ($V_g^{read}$) dependence of number of states.

In order to obtain long-term potentiation (LTP) and long-term depression (LTD) characteristics for neuromorphic applications, we have applied 1µs-wide pulse of increasing (0.2V to 8V with a step of 0.2V) and decreasing (-0.2V to -8V with a step of -0.2V) amplitudes to partially polarize the HZO stack in Fe-finFET (Figure. 2(a)). The drain voltage was kept at 0V during the programming operation. Each set (potentiation or depression) pulse was preceded by a reset pulse (±8V) to drive the conductance back to the starting point. The incrementing amplitude in each step increments the remnant polarization of HZO film by a small amount, which inevitably changes the threshold voltage and channel conductance of the Fe-finFET. Immediately after each programming pulse the READ operation of the memory state of the device was accomplished. During the READ operation the drain was kept at constant 100mV and a ramp voltage was applied at gate terminal. The ramp magnitude was varied during the read operations in a similar manner as [8] for eradicating the trapping in HZO film. The $G_{ch}$ was extracted from the $I_d$-$V_g$ curve of LTP and LTD operations at a specific gate voltage. We have used various gate voltage of values 0V, 0.25V, 0.5V, 0.75V, 1V and 1.25V to extract the $G_{ch}$.

Figure. 2(b) shows the LTP characteristics obtained by 1µs-wide increasing pulses and Figure. 2(c) shows the LTD characteristics obtained by pulses of the opposite polarity. We have applied a total of 26 positive and gradually increasing pulses for LTP and 26 negative and gradually decreasing pulses for LTD. Each pulse changes the remnant polarization of the HZO stack, which changes the threshold voltage ($V_{th}$) of the device, engendering a new memory state. Thus, we have obtained a total of 26 memory states excluding the initial states.

Figure. 2(d) shows the gradual change in $G_{ch}$ during LTP and LTD. The $G_{ch}$ is modeled according to one of our previous publication [8]. Although previous researchers have shown multilevel programming in HZO-based FE-FETs [9], this is the first to document the dependence of the number of conducting states on gate voltage for *READ* operation, $V_g^{read}$. Figure 2(e) shows that the number of available $G_{ch}$ states highly depends on $V_g^{read}$. Primarily, if $V_g^{read}$ is too low, most of the operational range will fall within the SUB-threshold region (below 200 nA) and do not count as distinct states. At the same time, however, the $G_{ch}^{ON}/G_{ch}^{OFF}$ ratio will be degraded when $V_g^{read}$ is too high. In the next section, we shall discuss about such trade-off for optimizing the system level performance.

### III. IMPLICATIONS TO NEURAL NETWORK OPERATION

To quantify the effect of these variations in Fe-finFETs, system-level neuromorphic simulations were performed [10]. We have adopted fully connected multilayer perceptron (MLP) model for the neural network architecture, where all the neurons of two consecutive layers are connected with each other. The network consists of an inputs layer with 784 nodes, one hidden layer with 200 nodes and an output layer with 10 nodes. The MNIST hand-written digit was initially cropped to 28×28 pixels, which was the input to the 784 input nodes. The details architecture of the neural network has been depicted in Figure 3(a). To quantify the effect of these variations in Fe-finFETs, system-level neuromorphic simulations were performed. We have adopted fully connected multilayer perceptron (MLP) model for the neural network architecture, where all the neurons of two consecutive layers are connected with each other. The network consists of an inputs layer with 784 nodes, one hidden layer with 200 nodes and an output layer with 10 nodes. The MNIST hand-written digit was initially cropped to 28×28 pixels, which was the input to the 784 input nodes. The details architecture of the neural network has been depicted in Figure 3(a). The training process in neural networkcan be divided into two categories, i.e., online and offline training. The online training requires constant update of the weights during the training and the online training necessitates high endurance of the synaptic devices. An endurance about $10^8$ cycles [7], makes our fabricated device appropriate for online training [9]. During offline training the weights are prefixed by software and the classification task is performed on the neural network. The training process involves two different tasks, i.e., feedforward and back propagation. The input from theneuron propagates in the forward direction (from input layer to output layer) and during this time multiply accumulate sum is performed on the synaptic array. The result obtained from the feed forward operation is then compared with the label or the correct answer to obtain the error in synaptic weights, which is propagated from the output to the input layer during back propagation operation. Once the training is complete, neural network performs inference operation, which involves only feed-forward operation, where synaptic weights are not changed. During the first scenario or online training 60000 MNIST image samples were used to train the hardware neural network. The back propagation algorithm with a batch size of 200 was adopted for training. In the online training scenario, D2D variation ($\Delta G_{ch}$) for each device was fixed at the beginning of training and remained unchanged throughout the training process. For C2C variation, $\Delta G_{ch}$ changed randomly during *READ* or *WRITE* operations throughout the training process for each device ineach cycle.The second scenario was offline training, in which neural network weight coefficients were pre-trained in software (without considering hardware related variation). The weights were subsequently written into hardware. Such a hardware neural network operates differently from software with errors because of D2D variation. In this case, weight adjustment of the neural network to compensate for hardware-related D2D variation was not possible.

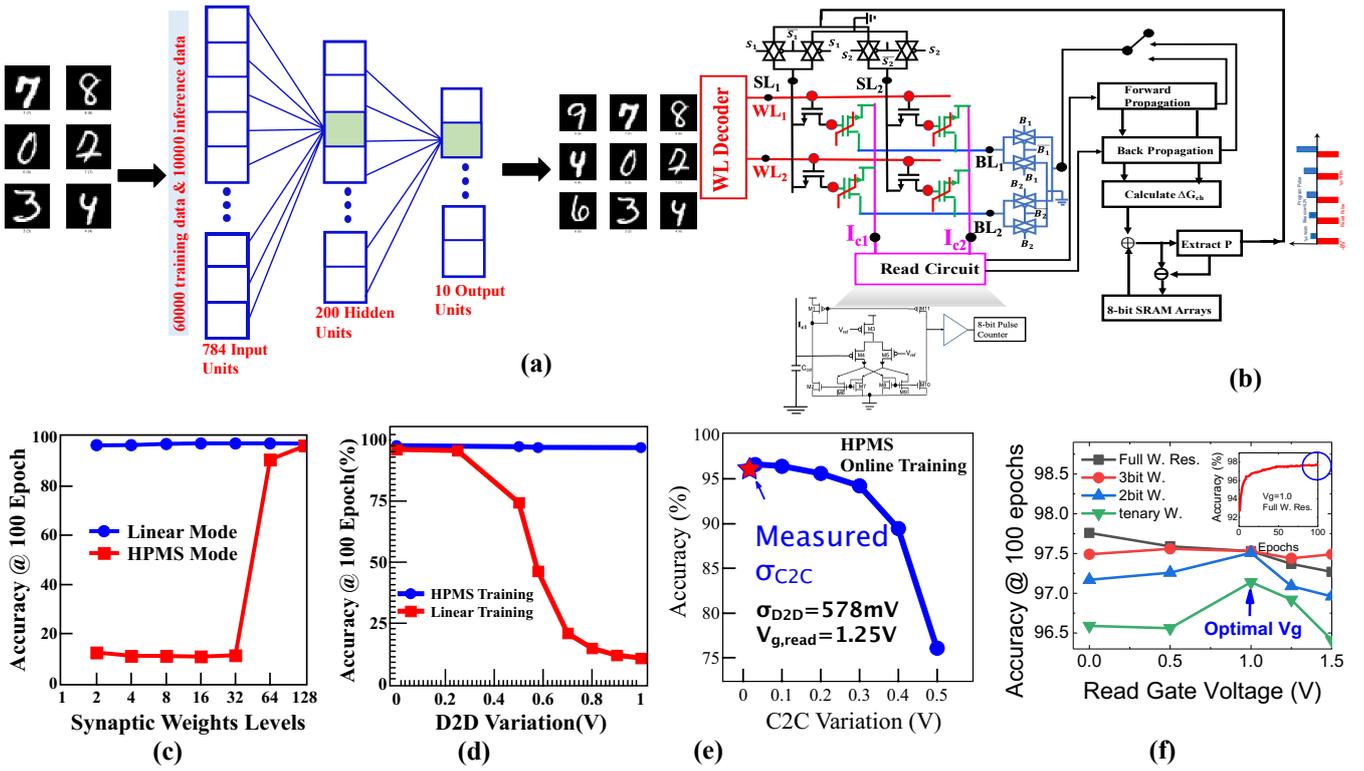

Fig.3. (a). MLP architecture considered during neuromorphic simulation in Cimulator platform. (b) The HPMS mode architecture used in Cimulator platform. The synaptic weights are stored in non-volatile Fe-finFET devices as channel conductance and the pseudo crossbar array of 1T-1Fe-finFET structure is used to perform MAC operation. The residual weight updates are stored and accumulated in a volatile digital SRAM memory cell. (c). HPMS mode shows excellent immunity towards low precision. (d). The HPMS mode shows excellent immunity towards device variations. (e). Negligible cycle to cycle variation induces very low errors during MAC operation. (f). Optimization of read voltage during MAC operation.

In this work we primarily focus on two different types of online weight update mechanism, where compensation of D2D variation by deploying peripheral circuit blocks are possible. The experimentally calibrated D2D variation and C2C variation exhibited a gaussian distribution with a standard deviation in the threshold voltage of $\sigma_{D2D}$=0.57833 V and $\sigma_{C2C}$= 0.0177 V, respectively. The modeled LTP and LTD characteristics with experimentally calibrated D2D and C2C variations or the standard deviation ($\sigma_{D2D}$, $\sigma_{C2C}$) in $V_{th}$ distribution was used to train the Fe-finFET-based multilevel perceptron (MLP)-based neural network with the MNIST dataset. The first one is trivial linear weight update mode and second one is the HPMS architecture [11], which has been exploited during the training operation to reduce the impact of device-to-device variations. The hardware platform consist of two different sections, the pseudo-crossbar array performing analog sum and a fully digital high precision accumulate and weight update section. The most critical task in hardware based neural network operation is multiply and accumulate sum, which is obtained by the vector matrix multiplication operation using Fe-finFET devices in our simulation framework. The synaptic weights are stored in terms of $G_{ch}$ of the devices. The applied programming voltage alters the synaptic weight or the $G_{ch}^{ij}$ during weight update or program erase operation. A read voltage pulse ($V_g^{read,i}$), equivalent to the input of the neuron, is applied to each column of the memory array which generates a vector matrix sum (current,$I_j$) to be detected by the current sense amplifier. The vector matrix sum can be given by $I_j=\sum G_{ch}^{ij} V_g^{read,i}$. The error coefficient is also sent to the HPMS weight update unit by current sum. Although a dedicated hardware acceleration unit for weight update can speed up the back-propagation process, the D2D variation among the Fe-finFETs hinders this process severely and it becomes more challenging to update the weights of each synapse precisely. We also evaluated the performance improvement in recognition accuracy obtained by adopting the HPMS weight update method to train the neural network. As discussed before, this methodology is similar to the $\Delta - \Sigma$ modulation method, which saves the weight residue for each training cycle to compensate for the insufficient weight resolution. Furthermore, the method exhibits considerable immunity to low precision and number of states. However, the method necessitates the use of additionalcircuits to store the residue portion of the weights, which arenot yet written to hardware. In HPMS mode the weight is rounded to the nearest quantization value with every residue discarded. The inference accuracy for the linear updatemode was approximately 10% (no recognition capability) until the weights had 6 bits or more (>90%accuracy). By contrast, the HPMS weight-update method requires considerably lower precision, and a 97% inference accuracy can be achieved with only two states or 1 bit (Figure. 3(c)). Therefore, in our case we update the weight of an only if the accumulated error coefficient exceeds a threshold value. The HPMS algorithm also displays excellent immunity towards D2D variation due to its ability to train at lower precision. Figure. 3(d) shows that the accuracy degradation was negligible due to D2D variation while the HPMS algorithm was deployed. Although in principle, C2C variation cannot be absorbed through HPMS weight update method, its effect is limited because of 17.7 mV variation. However, the degradation in performance in terms of accuracy becomes acute in the linear weight update scenario. The accuracy decreased from 97.46% to ~75% through D2D variation alone. The cumulative effect of the D2D and C2C variation reduced the recognition accuracy to 46.81%. Such nonlinear behavior is a distinct characteristics of neural networks. When variability is small within a certain threshold, cycle-to-cycle variation has a negligible effect. However, on reaching such threshold, the small randomness of 17.7 mV considerably degrades system accuracy. Figure 3(d) depicts how online HPMS training is useful for ensuring the robustness of neural networks in the presence of D2D variation. For example, with 10% high intrinsic maximum conductance for a given device, the HPMS training algorithm automatically adjusts the synaptic device to be trained to a higher $V_{th}$, which reduces the conductance by 10% to compensate for the variation. That is not the case for offline training or linear weight update method. However, online HPMS training cannot fully compensate for C2C variation (Figure 3(e)). Accuracy quickly degrades as C2C variation increase. Because a low read voltage and lower precision is a prerequisite for the low-power and application of the neural network, the HPMS weight update method is a promising algorithm to train Fe-finFET and does not require a high read voltage (more number of states).

Therefore, the impact of $V_g^{read}$ on the inference accuracy was investigated. Here, a $V_g^{read}$ of 1Vproved to be the optimal solution for all four cases displayed in the Figure.3(f). This read voltage dependency is the result of different on-off ratios of channel conductance at various gate voltages during read operations.

## IV. CONCLUSION

The overall performance of a neuromorphic system is the outcome of the confederated performance of the device, peripheral circuits, network architecture, and algorithm. In this study, we fabricated a nanoscale HZO FeFET device and analyzed the variability, especially for small devices We conclude that the effect of experimentally observed variations in system-level accuracy can be nullified by manipulating the neural network algorithm and architecture.

*Acknowledgement:* We are grateful to TSRI for providing us with advance CMOS fabrication platform, and NCHC for GPU computing facilities. This work was funded by Ministry of Science and Technology grants MOST-108-2634-F-006-008 and MOST-109-2628-E-492 -001 -MY3